\documentstyle[twoside,psfig]{article}
\newcommand{\be}{\begin{equation}}
\newcommand{\ee}{\end{equation}}
\newcommand{\ba}{\begin{eqnarray}}
\newcommand{\ea}{\end{eqnarray}}

\catcode`\@=11
\long\def\@makefntext#1{
\protect\noindent \hbox to 3.2pt {\hskip-.9pt
$^{{\eightrm\@thefnmark}}$\hfil}#1\hfill}       

\def\thefootnote{\fnsymbol{footnote}}
\def\@makefnmark{\hbox to 0pt{$^{\@thefnmark}$\hss}}    

\def\ps@myheadings{\let\@mkboth\@gobbletwo
\def\@oddhead{\hbox{}
\rightmark\hfil\eightrm\thepage}
\def\@oddfoot{}\def\@evenhead{\eightrm\thepage\hfil
\leftmark\hbox{}}\def\@evenfoot{}
\def\sectionmark##1{}\def\subsectionmark##1{}}

\oddsidemargin=\evensidemargin
\renewcommand{\thefootnote}{\fnsymbol{footnote}}

\newcounter{sectionc}\newcounter{subsectionc}\newcounter{subsubsectionc}
\renewcommand{\section}[1] {\vspace{12pt}\addtocounter{sectionc}{1}
\setcounter{subsectionc}{0}\setcounter{subsubsectionc}{0}\noindent
    {\tenbf\thesectionc. #1}\par\vspace{5pt}}
\renewcommand{\subsection}[1] {\vspace{12pt}\addtocounter{subsectionc}{1}
    \setcounter{subsubsectionc}{0}\noindent
    {\bf\thesectionc.\thesubsectionc. {\kern1pt \bfit #1}}\par\vspace{5pt}}
\renewcommand{\subsubsection}[1] {\vspace{12pt}\addtocounter{subsubsectionc}{1}
    \noindent{\tenrm\thesectionc.\thesubsectionc.\thesubsubsectionc.
    {\kern1pt \tenit #1}}\par\vspace{5pt}}
\newcommand{\nonumsection}[1] {\vspace{12pt}\noindent{\tenbf #1}
    \par\vspace{5pt}}

\newcounter{appendixc}
\newcounter{subappendixc}[appendixc]
\newcounter{subsubappendixc}[subappendixc]
\renewcommand{\thesubappendixc}{\Alph{appendixc}.\arabic{subappendixc}}
\renewcommand{\thesubsubappendixc}
    {\Alph{appendixc}.\arabic{subappendixc}.\arabic{subsubappendixc}}

\renewcommand{\appendix}[1] {\vspace{12pt}
        \refstepcounter{appendixc}
        \setcounter{figure}{0}
        \setcounter{table}{0}
        \setcounter{lemma}{0}
        \setcounter{theorem}{0}
        \setcounter{corollary}{0}
        \setcounter{definition}{0}
        \setcounter{equation}{0}
        \renewcommand{\thefigure}{\Alph{appendixc}.\arabic{figure}}
        \renewcommand{\thetable}{\Alph{appendixc}.\arabic{table}}
        \renewcommand{\theappendixc}{\Alph{appendixc}}
        \renewcommand{\thelemma}{\Alph{appendixc}.\arabic{lemma}}
        \renewcommand{\thetheorem}{\Alph{appendixc}.\arabic{theorem}}
        \renewcommand{\thedefinition}{\Alph{appendixc}.\arabic{definition}}
        \renewcommand{\thecorollary}{\Alph{appendixc}.\arabic{corollary}}
        \renewcommand{\theequation}{\Alph{appendixc}.\arabic{equation}}
        \noindent{\tenbf Appendix \theappendixc #1}\par\vspace{5pt}}
\newcommand{\subappendix}[1] {\vspace{12pt}
        \refstepcounter{subappendixc}
        \noindent{\bf Appendix \thesubappendixc. {\kern1pt \bfit #1}}
    \par\vspace{5pt}}
\newcommand{\subsubappendix}[1] {\vspace{12pt}
        \refstepcounter{subsubappendixc}
        \noindent{\rm Appendix \thesubsubappendixc. {\kern1pt \tenit #1}}
    \par\vspace{5pt}}

\newcommand{\textlineskip}{\baselineskip=13pt}
\newcommand{\smalllineskip}{\baselineskip=10pt}

\def\eightcirc{
\begin{picture}(0,0)
\put(4.4,1.8){\circle{6.5}}
\end{picture}}
\def\eightcopyright{\eightcirc\kern2.7pt\hbox{\eightrm c}}

\newcommand{\copyrightheading}[1]
    {\vspace*{-2.5cm}\smalllineskip{\flushleft
    {\footnotesize Modern Physics Letters A, #1}\\
    {\footnotesize $\eightcopyright$\, World Scientific Publishing
     Company}\\
     }}

\newcommand{\publisher}[2]{{\begin{center}\footnotesize\smalllineskip
    Received #1\\
    Revised #2
    \end{center}
    }}

\def\abstracts#1#2#3{{
    \centering{\begin{minipage}{4.5in}\footnotesize\baselineskip=10pt
    \parindent=0pt #1\par
    \parindent=15pt #2\par
    \parindent=15pt #3
    \end{minipage}}\par}}

\renewenvironment{thebibliography}[1]
    {\frenchspacing
     \ninerm\baselineskip=11pt
     \begin{list}{\arabic{enumi}.}
        {\usecounter{enumi}\setlength{\parsep}{0pt}
     \setlength{\leftmargin 12.7pt}{\rightmargin 0pt} 
         \setlength{\itemsep}{0pt} \settowidth
    {\labelwidth}{#1.}\sloppy}}{\end{list}}

\newcounter{itemlistc}
\newcounter{romanlistc}
\newcounter{alphlistc}
\newcounter{arabiclistc}

\newcommand{\fcaption}[1]{
        \refstepcounter{figure}
        \setbox\@tempboxa = \hbox{\footnotesize Fig.~\thefigure. #1}
        \ifdim \wd\@tempboxa > 5in
           {\begin{center}
        \parbox{5in}{\footnotesize\smalllineskip Fig.~\thefigure. #1}
            \end{center}}
        \else
             {\begin{center}
             {\footnotesize Fig.~\thefigure. #1}
              \end{center}}
        \fi}

\newcommand{\tcaption}[1]{
        \refstepcounter{table}
        \setbox\@tempboxa = \hbox{\footnotesize Table~\thetable. #1}
        \ifdim \wd\@tempboxa > 5in
           {\begin{center}
        \parbox{5in}{\footnotesize\smalllineskip Table~\thetable. #1}
            \end{center}}
        \else
             {\begin{center}
             {\footnotesize Table~\thetable. #1}
              \end{center}}
        \fi}

\def\@citex[#1]#2{\if@filesw\immediate\write\@auxout
    {\string\citation{#2}}\fi
\def\@citea{}\@cite{\@for\@citeb:=#2\do
    {\@citea\def\@citea{,}\@ifundefined
    {b@\@citeb}{{\bf ?}\@warning
    {Citation `\@citeb' on page \thepage \space undefined}}
    {\csname b@\@citeb\endcsname}}}{#1}}

\newif\if@cghi
\def\cite{\@cghitrue\@ifnextchar [{\@tempswatrue
    \@citex}{\@tempswafalse\@citex[]}}
\def\citelow{\@cghifalse\@ifnextchar [{\@tempswatrue
    \@citex}{\@tempswafalse\@citex[]}}
\def\@cite#1#2{{$\null^{#1}$\if@tempswa\typeout
    {IJCGA warning: optional citation argument
    ignored: `#2'} \fi}}

\def\pmb#1{\setbox0=\hbox{#1}
    \kern-.025em\copy0\kern-\wd0
    \kern.05em\copy0\kern-\wd0
    \kern-.025em\raise.0433em\box0}

\def\fnt#1#2{\footnotetext{\kern-.3em
    {$^{\mbox{\scriptsize #1}}$}{#2}}}

\def\fpage#1{\begingroup
\voffset=.3in
\thispagestyle{empty}\begin{table}[b]\centerline{\footnotesize #1}
    \end{table}\endgroup}

\def\runninghead#1#2{\pagestyle{myheadings}
\markboth{{\protect\footnotesize\it{\quad #1}}\hfill}
{\hfill{\protect\footnotesize\it{#2\quad}}}}
\font\tenrm=cmr10
\font\tenit=cmti10
\font\tenbf=cmbx10
\font\bfit=cmbxti10 at 10pt
\font\ninerm=cmr9

\font\eightrm=cmr8

\def\qed{\hbox{${\vcenter{\vbox{            
   \hrule height 0.4pt\hbox{\vrule width 0.4pt height 6pt
   \kern5pt\vrule width 0.4pt}\hrule height 0.4pt}}}$}}

\renewcommand{\thefootnote}{\fnsymbol{footnote}}    

\begin{document}
\setlength{\textheight}{7.7truein}  

\runninghead{HAMILTONIAN SOLUTION OF THE  SCHWINGER MODEL WITH COMPACT U(1)
$\ldots$}{HAMILTONIAN SOLUTION OF THE  SCHWINGER MODEL WITH COMPACT U(1)
$\ldots$}

\normalsize\textlineskip
\thispagestyle{empty}
\setcounter{page}{1}

\copyrightheading{}         

\vspace*{0.88truein}

\fpage{1}
\centerline{\bf EXACT SOLUTION OF THE  SCHWINGER MODEL }
\baselineskip=13pt
\centerline{\bf WITH COMPACT U(1)}
\vspace*{0.37truein}
\centerline{\footnotesize ROM\'AN LINARES, LUIS F. URRUTIA and J. DAVID
VERGARA}
\baselineskip=12pt
\centerline{\footnotesize\it Departamento de F\'\i sica de Altas Energ\'\i as,
Instituto de Ciencias Nucleares}
\baselineskip=10pt
\centerline{\footnotesize\it Universidad Nacional
Aut\'onoma de M\'exico}
\baselineskip=10pt
\centerline{\footnotesize\it Apartado Postal  70-543, 04510,
M\'exico D.F., M\'exico}
\vspace*{10pt}

\vspace*{0.225truein}

\publisher{(received date)}{(revised date)}

\vspace*{0.21truein}
\abstracts{The exact solution of the  Schwinger model with compact gauge
group $U(1)$ is presented. The compactification is
imposed by demanding that the only surviving true
electromagnetic  degree of freedom has angular character. Not surprinsingly, this
topological condition
defines a  version of the Schwinger
model which is different from the standard one, where $c$ takes values on the line.
The main consequences are: the spectra of the zero modes is not degenerated and
does not correspond to the equally spaced
harmonic oscillator,  both the electric charge and a modified gauge
invariant chiral charge are conserved (nevertheless, the axial-current anomaly
is still present) and, finally, there is no need to
introduce a $\theta$-vacuum. A comparison with the results of the standard
Schwinger model is  pointed out along the text.}{}{}



\vspace*{1pt}\textlineskip  
\section{Introduction}  
\vspace*{-0.5pt}
\noindent

Using the Hamiltonian approach we solve the Schwinger model, with
compact gauge group $U(1)$, which we call the compact Schwinger model (CSM) in
the sequel. The standard Schwinger  model
\cite{Schwinger} has been solved in many ways and we have not attempted here
to provide a complete list of all the related references
\cite{general,ADAMS,Capri,Manton,HetHo,Shifman,Link,IsoMurayama,HallinLiljenberg}.
We will refer to the latter as the non-compact Schwinger model (NCSM).

The
compactification  of the gauge group $U(1)$ is realized by demanding that
the only surviving electromagnetic degree of freedom, called $c$ in the sequel,
behaves as an angular variable living in a circle of
lenght $\frac{2\,\pi}{e\, L}$ , i.e. $-\,\pi/eL \leq c \leq +\,\pi/eL$.
Not surprinsingly, the compactification prescription leads to a model which
drastically differs from the NCSM, as will be seen along the text. 

Many solutions of the NCSM, where the electromagnetic degree of freedom $c$
lives in the line $\{-\infty, +\infty\}$, start from
considering $c$ as an angular variable. Nevertheless, using apropriate boundary
conditions, the corresponding authors manage to unfold the circle into the line,
i.e. to go from $U(1)$ to its
universal covering \cite{Manton,Shifman,Link,IsoMurayama}.

Here we maintain the angular character of $c$ and fully explore the consequences
of this choice. It is important to emphasize that our results follow uniquely from
the compactification
condition, together with the standard definitions of both a scalar product and
the hermiticity requirements in the corresponding Hilbert space.

A partial solution of the CSM was found in Ref. \cite{GMVU},
using the loop approach to this problem \cite{GP}, and served as a
motivation for the work presented here. These partial results coincide
with those obtained in this work. Previous progress towards the complete solution of
the CSM were
reported in \cite{PREREP}.

Since both models, the CSM and the NCSM, differ only in the topology of
the gauge group, it is neither surprising  that the Hamiltonian method
employed in the solution of the latter would be also effective for
the former. For this reason and with the necessary
modifications, we relay in the work of Refs. \cite{Link,IsoMurayama} which
provide a complete  Hamiltonian solution of  the NCSM.

\setcounter{footnote}{0}
\renewcommand{\thefootnote}{\alph{footnote}}

\section{CSM, gauge invariant degrees of freedom and commutator algebra}
\noindent
The model is described by the Lagrangian
\begin{equation}
{\cal L} = -\frac{1}{4} F_{\mu\nu}F^{\mu\nu}
           +\bar{\psi} \gamma^{\mu} \left(i\partial_{\mu}
           - e A_{\mu}\right)\psi
          \label{LAG}
\end{equation}
where $F_{\mu\nu}=\partial_{\mu}A_{\nu} -\partial_{\nu}A_{\mu}$,
 $\bar{\psi}=\psi^\dagger \gamma^0$ is a  Grassmann valued
fermionic field and  we are using units such that $\hbar=c=1$.
We consider the coordinate space  to be $S^1$ and  we will require
periodic$\,$(antiperiodic)
boundary conditions for the fields $A_{\mu}(x)\,\, (\psi (x))$,
where  $L$ is  the length of the circle.
The gamma matrices are:
$\gamma^0=\sigma_1,\; \gamma^1=i\sigma_2,\; \gamma^5=-\gamma^0\gamma^1=
\sigma_3$, where $\sigma_i$ are the standard Pauli matrices.
We use the  signature $(+,-),\;\; i.e.\;\; \eta_{00}=-\eta_{11}=1$.


After the standard canonical analysis of the Lagrangian density
(\ref{LAG}), describing the configuration space variables
$A_0,\,  A_1$ and $\psi$, we obtain
\begin{equation}
{\cal H} = \frac{1}{2} E^2
           + i \psi^\dagger\sigma_3
           \left(\partial_1 + i e A_1\right)\psi
              - A_0\,\left(\partial_1\, E - e\, \psi^\dagger \,
 \psi \right), \qquad \Pi_0\approx 0,
\label{HAMDEN}
\end{equation}
where the  corresponding  canonical momenta are
$\Pi_0,\, \Pi_1=F_{01}=E $ and $\Pi_\psi=-i\, \psi^*$. Conservation
in time of the primary constraint $\Pi_0\approx 0$
leads to the the Gauss law constraint
\begin{equation}\label{GL}
{\cal G}=\partial_1 E - e \psi^\dagger\psi\approx 0.
\end{equation}
There are no additional constraints. At this stage we partially fix the gauge in
the electromagnetic potential by choosing $A_0=0, \,\,  \Pi_0=0$.
The only remaining constraint ${\cal G}$ is first class and
it will be imposed strongly upon the physical states of the system.

From now on we use the notation $A_1=A$ for the surviving
electromagnetic degree of freedom. Also we have
$\psi = (\psi_+,\psi_-)^\top$, where
$\top$ denotes transposition.  Applying the standard canonical quantization procedure to the
resulting Poisson brackets algebra, we obtain the well known commutator (anticommutator)
algebra for the involved fields.


The  previous choice of gauge does not completely fix the electromagnetic
degrees of freedom, leaving  the Lagrangian density (\ref{LAG}) still invariant
under the gauge transformations
\begin{equation}
  \psi \rightarrow  {\rm e}^{ie \alpha(x)} \psi,
  \quad A_\mu \rightarrow A_\mu - \partial_\mu \alpha (x), \label{GGT}
\end{equation}
generated by $\cal G$. The constant piece $\alpha_0$ of
the function $\alpha(x)$ is
irrelevant in the above transformation, leading to an overall phase in the fermionic field.
In the sequel we consider ${ \bar \alpha}(x)= \alpha(x) - \alpha_0$
as the function generating the gauge transformations.
As it is well known, there are two families of gauge transformations:(1) those continuously
connected to the identity,
called small gauge transformations (SGT), characterized by  the  functions
\begin{equation}
{\bar \alpha}_S(x) = b \left(e^{i 2 \pi n x  / L} -1  \right),\quad
{\bar \alpha}_S(0) ={\bar \alpha}_S(L) = 0,
\end{equation}
which are periodic in $x$. (2) The second family corresponds to the so called large gauge
transformations (LGT), which are generated  by the
non-periodic functions
\begin{equation}
\label{LGT}
  {\bar \alpha}_L(x)={2 \pi
n \over e L}x=2\,n {\bar c}\,x , \quad n=\pm 1, \pm 2, \dots , \quad {\bar c}={ \pi
\over e L}, \quad {\bar \alpha}_L(0)=0 .
\end{equation}
Let us emphasize that in both cases we have
\be
\label{ALFCOND}
{\bar \alpha}(0)=0.
\ee
At this stage we {\bf define} the CSM by demanding that
the only true degree of freedom arising from the electromagnetic
potential in one dimension, which is the zero mode
$c$, be restricted
to the interval
\begin{equation}\label{COMPACT}
- {\bar c}\leq c=\frac{1}{L}\int_0^L A(z)\, dz \leq {\bar c}.
\end{equation}
The compactification condition (\ref{COMPACT})  implies that two values of $ c $
differing by
$2\,{\bar c}\, N= \frac{2\, \pi\, N}{e L}$
must be  identified, corresponding to one point in such configuration space.

Next we show that the basic degrees of freedom in the CSM are in fact fully gauge
invariant.
Let us consider the following Fourier decomposition for the
electromagnetic potential $A$, the field strength $E$
and the gauge transformation function ${\bar \alpha }$
\begin{eqnarray}\label{RFD}
 A(x,t) &=& c(t) + \sum_{m \neq 0 } A_m(t) \ { \rm e}^{\frac{2 \pi  i m
}{L} x},
\quad
E(x,t) = E_0(t)+ \sum_{m \neq 0} E_m(t) \ { \rm e}^{- \frac{2 \pi  i m }{L} x},
\nonumber \\
 {\bar \alpha}(x) &=& \sum_{m \neq 0}
{\bar \alpha}_m \ { \rm e}^{\frac{2 \pi  i m }{L} x},\qquad {\rm where}\qquad  0={\bar \alpha}(0) =  \sum_{m \neq 0} {\bar \alpha}_m.
\end{eqnarray}

Under a general gauge transformation $ A(x) \rightarrow A(x) -
\frac{\partial {\bar  \alpha}(x)}{\partial x} $,
the corresponding modes change as
\begin{equation}\label{GTOM}
c \rightarrow c -\frac{1}{L} ({\bar \alpha} (L) - {\bar \alpha}(0)) ,
\quad A_m \rightarrow A_m -
 \frac{2 \pi i m}{L} \ {\bar \alpha}_m, \quad m\neq 0 .
\end{equation}
Clearly, the zero mode $c$ is invariant under SGT. As for  LGT, $c \rightarrow c - \frac{2
\pi n}{e L
}$, but these points
must be identified according to the compactification condition
(\ref{COMPACT}). In other words, $c$ is also invariant under LGT. Summarizing,
the zero mode $c$ is fully gauge
invariant. This is a consequence of our choice of topology for $c$ and  provides
the fundamental difference with
the NCSM, leading to  all the remaining non-standard features of the CSM.

Next we consider  the
expansion of the fermionic
variables  in a background
electromagnetic field $A(x)$. According to Ref. \cite{IsoMurayama}, these
can be written as
\begin{equation}
\psi_+(x,t)=\sum_{n} a_n \phi_n(x){\rm e}^{-i\epsilon_n t}, \quad
\psi_-(x,t)=\sum_{n} b_n \phi_n(x)
{\rm e}^{i\epsilon_n t},\label{psi12}
\end{equation}
where  $a_n, b_n$  are independent fermionic  annihilation
operators satisfying the  standard  anticommutators.
The states
$\psi_+ \ (\psi_-)$ describe the positive \ (negative) chiral (eigenvalues of
$\gamma_5$)    sectors of
the model. The  functions $\phi_n$ , together with the
energy eigenvalues $\epsilon_n$ are given by
\begin{equation}
\phi_n(x)={1\over\sqrt L}e^{-i \epsilon_nx -ie\int_0^x
A(z)dz},\quad
\epsilon_n=
{2\pi\over L}\left(n+{1\over2} -{eL\over 2\pi}c\right).
\label{VFP}
\end{equation}
Rewriting the fermionic sector  of the Hamiltonian density
(\ref{HAMDEN}) as
${\cal H}_F=
\psi^\dagger h_F \psi$, we observe that  the corresponding
eigenvalues of $h_F$
are
$+ \epsilon_n$ and $- \epsilon_n$ for the positive and negative quirality
sectors, respectively.

Since $c$ is invariant under both LGT and SGT, each
energy eigenvalue $\epsilon_n$ is fully gauge invariant. Furthermore, according
to the definition
(\ref{VFP}), we obtain
\be
\phi_n \to {\rm e}^{ie\bar \alpha(x)}\phi_n,
\ee
under gauge transformations. Here
we have used the condition $\bar \alpha(0)=0$, which is valid for both LGT and SGT. As
a consequence of the above
properties and in order to recover the trasformation law (\ref{GGT}) of the fermionic
field $\psi$, we are led to
\begin{equation}\label{GTAYB}
a_n \rightarrow a_n, \quad
b_n \rightarrow  b_n.
\end{equation}
for the gauge transformation of the fermionic operators $a_n$ and $b_n$.

In other words, consistency among the compactification condition (\ref{COMPACT}), the
transformation law (\ref{GGT}) and the definition (\ref{VFP}) for $\phi_n$ demands
that the basic
fermionic operators $a_n$ and $b_n$ are fully gauge invariant in the compact case.
Let us emphasize that the above property establishes the main difference  between the
CSM and the NCSM. 

Following the same steps in
the NCSM we obtain that  the change $c \rightarrow c - \frac{2
\pi n}{e L }$, under LGT and with these points not identified,  implies that the
individual energy eigenvalues are
not gauge invariant, i.e.  $\epsilon_n \to \epsilon_{n+1}$. This leads to
$\phi_n(x)  \to {\rm e}^{-ie \bar \alpha(x)} \phi_{n+1}$. Again,
in order to satisfy the transformation property (\ref{GGT}) of the fermionic field,
we must have now that $a_n\rightarrow a_{n+1}, b_n \rightarrow b_{n+1} $, under LGT,
which is the well known result in the NCSM.

In this way, it is transparent that the topological behavior of $c$, i.e. compact
versus non-compact case, implies  completely different transformation laws for the
operators $c, a_{n}$ and $b_n$ under gauge transformations.

Using the Fourier expansions (\ref{RFD}) and (\ref{psi12}), we rewrite the commutators
for the fields in terms of the correspondig modes.
In particular, the commutator $[ E(x), \psi_\alpha(y)]=0$ leads to
\begin{equation}\label{CREa}
[E_m, a_n] = \frac{ie}{2 \pi m }\left( a_n - a_{n+m}  \right),
\ \  m \neq0, \qquad [E_0, a_n]=0,
\end{equation}
with analogous relation for $[E_m, b_n]$.
The remaining commutators are
\begin{equation}\label{RCOM}
[ A_k, A_l]=0= [ E_k, E_l], \quad  [ A_k, E_l]= \frac{i}{L} \delta_{kl},
\quad [ A_k, a_n]=0, \quad [ A_k, b_m]=0.
\end{equation}
Next, we concentrate on the commutator algebra of the Fourier modes. To this end, it
is convenient to
introduce the following  operators
\begin{equation}\label{DC}
  j_{++}^{nm}=a_n^\dagger a_m, \hspace{1cm}
  j_{--}^{nm}=b_n^\dagger b_m, \hspace{1cm}
  j_{+-}^{nm}=a_n^\dagger b_m, \hspace{1cm}
  j_{-+}^{nm}=b_n^\dagger a_m.
\end{equation}
 Another useful combinations of the above
fermionic operators are the currents
\begin{eqnarray}\label{JMMX}
&&j_\pm(x)= \psi_\pm{}^\dagger(x) \psi_\pm(x)= \frac{1}{L}
\sum_{n=-\infty}^{+ \infty} {\rm e}^{\mp \frac{2 \pi  i n}{L} x}
\ j_\pm{}^n,
\end{eqnarray}
where
\begin{equation}
\label{JMM}
j_+{}^n =\sum_{m=-\infty}^{\infty} j_{++}^{m,m+n}, \qquad
j_-{}^n =\sum_{m=-\infty}^{\infty} j_{--}^{m+n,m}.
\end{equation}

At this stage we introduce the $\zeta-$regularized form
of the currents defined in Eq.(\ref{JMM}),
\begin{equation}
\label{REGJ}
j_+{}^n|_{\rm reg}={\rm lim}_{s\rightarrow 0}
\sum_{m=-\infty}^{\infty} \frac{1}{\lambda_{m,s}}a^\dagger_{m} a_{m+n}, \quad
j_-{}^n|_{\rm reg}={\rm lim}_{s\rightarrow 0}\sum_{m=-\infty}^{\infty}
\frac{1}{\lambda_{m,s}}b^\dagger_{m+n } b_{m},
\end{equation}
where the regulator is given by $\lambda_{n,s}= |\lambda\,
\epsilon_n|^s$, with
$\lambda$ been a parameter with dimensions
of inverse energy. 
In the sequel we will drop the subindex $ |_{\rm reg}$ from
the above
currents, but we will always consider their form (\ref{REGJ})
in calculating any relation involving
them. At
the end of the calculation we will take the $s\rightarrow 0$
limit.
In other words, we will construct an algebra among regularized
objects, which will be further restricted to the action upon the
physical
Hilbert space of the problem. It can be shown that the regularized current algebra
of the operators (\ref{REGJ}) is given by
\begin{equation}\label{CRROFJ}
 [ j_+{}^n,   (j_+{}^m)^\dagger ]= n \delta_{m,n}, \qquad
[ j_-{}^n,  ( j_-{}^m)^\dagger ]= n \delta_{m,n}, \qquad
[ j_+{}^n,   j_-{}^m ]=0,
\end{equation}
together with the hermiticity properties
$(j_\pm{}^m)^\dagger = j_\pm{}^{-m}$.
The  above commutation relations are the same as those obtained in the NCSM
\cite{Link,IsoMurayama}.

In order to satisfy the commutation relations (\ref{CREa})
and (\ref{RCOM}) we make the ansatz
\begin{equation}\label{AFORE}
 E_m=\frac{1}{i L}\frac{\partial}{\partial A_m}  - \frac{e}{2 \pi  i m}
 \left( j_+{}^m + (j_-{}^m)^\dagger \right),\ \ m\neq 0, \quad E_0=
 \frac{1}{i L} \frac{\partial}{\partial c },
\end{equation}
which clearly satisfies the third commutation relation in (\ref{RCOM}). 
 Substituting the  expressions
(\ref{AFORE})
in the corresponding commutators of Eq.(\ref{CREa}) we obtain
\begin{equation}\label{DAODA}
\frac{\partial a_n}{\partial A_m}= - \frac{e L}{ 2 \pi m} a_n, \
m \neq 0,\quad \frac{\partial a_n}{\partial c}=0.
\end{equation}
The above equation leads to the following solution for the fermionic
operators with respect to their dependence on the gauge field
\begin{equation}\label{SOLFO}
 a_m= {\rm exp} \left( -\frac{e L}{2 \pi }
\sum_{k\neq0} \frac{1}{k} A_k \right) {\bar a}_m, 
\end{equation}
where ${\bar a}_m$ are new fermionic operators which are independent
of the gauge field $A_k$
and which also  satisfy the basic fermionic  anticommutation relations.
The  expression (\ref{SOLFO}) reproduces the fully gauge invariant character
of $a_n$. In fact, under the gauge transformation (\ref{GTOM}),
the exponential in (\ref{SOLFO})  changes by a factor
${\rm exp}( i e \sum_{k \neq 0} {\bar \alpha}_k) $ which is exactly
${\rm exp}( i e {\bar \alpha}(0))=1$, according to the fourth relation in 
(\ref{RFD}). Analogous results are  obtained for the operators $b_n$.

With the ansatz (\ref{AFORE}), the Gauss law
${\cal G}(x) =- \frac{1}{L}\sum {\rm exp} (-\frac{2 \pi i mx}{L})
{\cal G}_m $ reduces to
\begin{eqnarray}
\label{GAUSSL}
{\cal G}_0&=& e \left( j_+{}^0 + (j_-{}^0)^\dagger \right) =e Q
\nonumber \\
{\cal G}_m &=& 2 \pi i m \ E_m + e \left( j_+{}^m +
(j_-{}^m)^\dagger \right)=\frac{2 \pi m}{L} \frac{\partial}{\partial A_m}, \quad
m\neq 0.
\end{eqnarray}
Our expression  (\ref{GAUSSL}) for the Gauss law constraint
is somewhat different from the one obtained in  Ref. \cite{IsoMurayama}.

Sumarizing, we have shown that the compactification condition (\ref{COMPACT})
implies that  the basic operators which are used to solve the model:  $c$, $a_n$ and
$b_n$ are fully gauge invariant. Also, the Gauss constraint  implies that
the wave function of the system must have zero electric charge, being independent of the electromagnetic modes
$A_m,\ m\neq 0$.

\section{Fermionic  Fock space}
\noindent
We now construct the fermionic
Fock space in a background electromagnetic field. Starting from the
vacuum
$|0 \rangle$ annihilated by the operators $a_n$ (positive quirality
sector),
$b_n$  (negative quirality sector),
the Dirac vacuum $|vac \rangle$
is constructed  in such a way that all negative energy levels are
filled. Our compactification condition  (\ref{COMPACT}) for the
electromagnetic variable $c$  implies that  all  levels with $n \leq -1
\ ( n\geq 0
) $ have negative energies for the
 positive  \ (negative) quirality sectors, respectively.  In this
way, the Dirac vacuum is
\begin{equation}
\label{DVD}
  |vac \rangle =
                 \prod_{n=-\infty}^{-1}a_n^\dagger |0 \rangle \otimes
                 \prod_{n=0}^{\infty}b_n^\dagger |0 \rangle.
\end{equation}
Using the  $\zeta-$regularized  expresions for the operators electric charge $Q$,  chiral charge $Q_5$ and
energy $H_F$ \cite{IsoMurayama}, we obtain the following
eigenvalues on  the Dirac
vacuum state 
\begin{eqnarray}
  Q |vac\rangle &=& 0, \qquad   Q_5 |vac\rangle = -\frac{ecL}{\pi } |vac \rangle ,
\nonumber \\
  H_F |vac\rangle &=& \frac{2\pi}{L}\left \{ \left( \frac{ecL}{2\pi}\right)^2
-\frac{1}{12} \right \}
 |vac\rangle \equiv {\varepsilon}_0 |vac\rangle,
\label{EDV}
\end{eqnarray}
At this level it is already convenient to introduce the
modified chiral charge
\begin{equation}
\label{MCC}
{\bar Q}_5= lim_{s\rightarrow0}\sum_{n= -\infty}^{+ \infty}
 \frac{1}{|\lambda \epsilon_n|^s}\left( { a}_n^\dagger { a}_n
- { b}_n^\dagger { b}_n  \right)+\frac{ecL}{\pi},
\end{equation}
which is fully gauge invariant
in the CSM. This is not the case in NCSM, where ${\bar Q}_5 \rightarrow
{\bar Q}_5+ 2n$  under LGT. The eigenvalues of ${\bar Q}_5$ are even
numbers in the fermionic Hilbert space. In the sequel we will refer
to ${\bar Q}_5$ as the modified chiral charge of the system. Also, a given
 eigenvalue $2\, M$ of
${\bar Q}_5$ will be referred to as the $M$-chiral sector of the theory.
In this way, the
second equation (\ref{EDV}) reads ${\bar Q}_5 |vac\rangle=0$, which
assigns zero modified chiral charge to the Dirac vacuum. Without writting  the
explicit label of zero electric charge, which will be undestood for all
physical states in the sequel,  we denote
\be
|vac\rangle= \left | {\varepsilon}_0, 0 \right \rangle.
\ee

It can be shown that the  current operators $j_+^n$ and
$j_-^n$ do
not change either the electric or  the chiral charge. Also, the action of  any
linear
combination of the operators $j_{+-}^{pq}\,\, (j_{-+}^{pq})$, which leaves invariant
the electric charge, will change the chiral charge by $+\,2\, (-\, 2)$ units.

Next, using the operators
(\ref{DC}) we can construct additional states with minimum energy, but
different chirality. The are
\begin{equation}
\label{VED}
  \left |{\varepsilon}_N,2N \right \rangle =
                 \prod_{n=-\infty}^{N-1}a_n^\dagger |0 \rangle \otimes
                 \prod_{m=N}^{\infty}b_m^\dagger |0 \rangle, \quad 
{\varepsilon}_N(c)=\frac{2 \pi}{L}\left\{\left( N- \frac{ecL}{2 \pi}
\right)^2-\frac{1}{12}\right\}.
\end{equation}
All the states in (\ref{VED}) have zero electric charge. They  satisfy
the recursions
\begin{eqnarray}
 j_{+-}^{NN}\left |{\varepsilon}_N, \, 2N \right \rangle =
                     \left | {\varepsilon}_{N+1},\, 2(N+1) \right \rangle,
\quad j_{-+}^{N-1N-1}\left |{\varepsilon}_N,\, 2N \right \rangle =
                     \left |{\varepsilon}_{N-1},\, 2(N-1) \right \rangle.
\label{COV}
\end{eqnarray}

Summarizing, from the Dirac  vacuum we have so far  constructed states with
minimum energy for each
possible chirality. Each one of these states can be considered as a
{\it local vacuum} in the corresponding  quirality sector.

The properties
\begin{equation}
\label{CHFJ}
j_\pm{}^n \left |{\varepsilon}_N,2N \right \rangle =0,\quad n\geq 1,\qquad
\left[H_F, (j_{\pm}^n)^\dagger \right]= \frac{2 \pi n}{L} \ (j_{\pm}^n)^\dagger.
\end{equation}
allow us to construct
the complete fermionic Fock space in the background electromagnetic field. It
will consist
of all the local vacuums  (\ref{VED}), together with all  possible
states constructed from them by the application of an arbitrary number
of the current operators ${(j_\pm{}^n })^\dagger, n=1,2, \dots$ defined in
Eq. (\ref{REGJ}).
The  spectrum of this  fermionic Fock space is
\be
\{ {\varepsilon}_N + \frac{2 \pi}{L} M, M=1,2, \dots, ,\ N=0, \pm1, \pm 2, \dots  \}.
\ee
In this way, the fermionic Hamiltonian in the external field can be rewritten
 in the Sugawara form \cite{SUGA}
\begin{equation}
\label{HAMCUAD}
H_F= {\varepsilon}_N(c)+ \frac{2\pi}{L} \sum_{n>0} \left((j_+^n)^\dagger  j_+^n +
(j_-^n)^\dagger  j_-^n\right).
\end{equation}

\section{Complete solution and comments}
\noindent
The next step  is to write  the complete  Hamiltonian
$H=H_{EM}+H_{F}$ in terms the currents operators,
together with
the electromagnetic degrees of freedom, which are   the zero
mode of the electric
field $\partial /\partial c$, and the zero  mode of the gauge potential
$c$. The result is
\begin{equation}\label{HM}
 H=H_{EM}+H_{F}=H_0+\sum _{n>0}H_n-\frac{2\pi}{12 L},
\end{equation}
where
 \begin{eqnarray}\label{HME}
  H_0&=&\frac{\pi}{2 L} \left ( Q^2+\left({\bar Q}_5-
\frac{e c L}{\pi}\right)^2 \right) -\frac{1}{2L}
 \left(\frac{\partial}{\partial c} \right)^2,
\nonumber \\
  H_n&=& \frac{2\pi}{L}
  ((j_+^n)^\dagger  j_+^n + (j_-^n)^\dagger  j_-^n ) +
  \frac{e^2L}{4 \pi^2 n^2}
  ((j_+^n)^\dagger + j_-^n) (j_+^n+(j_-^n)^\dagger).
\end{eqnarray}
Following Refs. \cite{Link,IsoMurayama}, we have explicitly used  the
Gauss law constraint (\ref{GAUSSL}) to express the electric field modes $E_m$ in terms
of the fermionic currents.

In order to diagonalize the expression (\ref{HME})  for the
Hamiltonian, we use the  Bogoliubov transformations ${\tilde j}_+ ^n= U_n^\dagger\,
(j_+ ^n )\,U_n$, given in \cite{Link,IsoMurayama},  for the currents
operators. The Bogoliubov transformation
affects  only  the   modes $n\geq 1$  of the system and, in particular, the
currents $j_{\pm}^0$, or equivalently $Q$ and ${\bar Q}_5$,  remain unchanged.
In this way,  the fully  rotated  Hamiltonian
$
{ H}_B=U^\dagger \,H( { j}_+^n, { j}_-^n) \, U= H( { {\tilde j}}_+^n, { {\tilde j}}_-^n),
$
 is
 \begin{eqnarray}
\label{HTOTAL}
 { H}_B= \frac{\pi}{2L} \left(Q^2+\left({\bar Q}_5-
\frac{e c L}{\pi}\right)^2 \right)
      -\frac{1}{2L} \left(\frac{\partial}{\partial c} \right)^2
+ \sum_{n>0} \frac{{\cal E}_n}{n}\left(({ j}_+^n)^\dagger
{ j}_+^n + ({ j}_-^n)^\dagger { j}_-^n \right),
\end{eqnarray}
up to an infinite constant.

The general structure of the states in the full Hilbert space of  the model will be
of the type
\begin{equation}
\label{GENFORM}
|\hbox{state} \rangle= F(c) \times |\hbox{fermionic}\rangle.
\end{equation}
The whole wave function will have zero electric charge and definite
chiral charge, which
really implies a condition only upon de fermionic
piece. The  strategy to construct the Hilbert space  will be to start
from the zero modes $ F_N(c) \times  \left| {\varepsilon}_N,2N \right\rangle $
and to subsequently apply
all possible combinations of the
raising operators $({ j}_\pm{}^m)^\dagger$. 

First we consider the zero modes.
They correspond to the case of zero fermionic excitations above the corresponding Dirac
vacuum and can be written as
\begin{equation}
\label{ZERMOD}
{ |N \rangle}_B= F_N(c) \times \left| {\cal E}_N,2N \right\rangle.
\end{equation}
The subscript $B$ in any ket is to remind us that such vector is written in the
Bogoliubov rotated
frame, where the Hamiltonian has the form (\ref{HTOTAL}).
Its action  upon the above wave functions reduces to the
following Schroedinger equation for the zero mode wave functions
$F_N(c)$
\begin{equation}
\label{ZEROM}
\left(-\frac{1}{2L}\left( \frac{\partial}{\partial c} \right )^2
+ \frac{e^2L}{2\pi}\left( \frac{2\pi N}{eL} -c \right )^ 2
\right)F_N(c)= {\cal E}_{N,0} F_N(c),
\end{equation}
which corresponds to a  piecewise harmonic oscillator.
Each sector, labeled by $N$,  is defined in the interval $-{\bar c}\leq c \leq
{\bar c}$.

For arbitrary functions $F(c)$ and $G(c)$,  we define the inner product in the
standard way  as
\begin{equation}
\label{scalarp}
\langle F | G \rangle = \int^{\bar c}_{-\bar c} dc\ F^*(c) G(c).
\end{equation}
Next we demand the hermiticity of the zero mode electric field operator
$E_0=\frac{1}{i\,L}\,
\frac{\partial}{\partial\,c}$, together with the Hamiltonian (\ref{ZEROM}). The above requirements lead to the boundary conditions
\begin{equation}
\label{BCWF}
F_N|_{c=-{\bar c}}=F_N|_{c=+{\bar c}},
\qquad {\partial F_N \over \partial c}|_{c=- {\bar c}}=
{\partial F_N \over \partial c}|_{c=+ {\bar c}},
\ee
for the wave function $F_N$ and their derivatives. 

A fundamental difference  between the CSM and the
NCSM arises in the energy spectrum $\{{\cal E}_{\alpha, N, 0}\}$ of  zero
mode sector. Here $\alpha=0,1,2, \dots $ labels the eigenvalues of the zero mode $0$ in the
$N$-chiral  sector of the model.
The solution corresponding to $N=0$ has been already discussed in
Ref.\cite{GMVU},
together with the corresponding wave functions. Here we extend the
calculation
for arbitrary $N\neq 0$. The general solution of the above Schroedinger equation can be expressed
in terms
of cylindrical parabolic functions \cite{Abramowitz}. The energy eigenvalues are parametrized
as ${\cal E}_{\alpha, N,0}=-\frac{e}{\sqrt \pi}\, a_{\alpha, N}$ , where  $a_{\alpha, N}$
is determined by a complicated trascendental equation arising from the boundary conditions \cite{PREREP}.
As in the $N=0$ case \cite{GMVU},
this function can
only be determined numerically for arbitrary $l=\frac{e\,L}{{\pi}^{3/{2}}}$.
In Figs.1 and  2  we show
the results for $a_{\alpha, N}$ versus $l$, for the choices $\alpha = 0,1,2$ and $N=0, 1, 2, 3$.

Among the zero modes,  we now focus on  the local minimum
($\alpha=0$) energy  states:
${|0,\, N, \, 0  \rangle}_B$, for each  chiral sector $N$ of the theory. They have energies
${\cal E}_{0,\, N,\, 0}$.
An important consequence of the compactification prescription is that these states
are not fully degenerated as they are in the NCSM. Most importantly, from  the numerical calculation we
find that the absolute minimum value of
${\cal E}_{0,\, N,\, 0}$ correspond to $N=0$. Thus, in the compact case the physical,
non-degenerated, vacuum of the theory is $|0,\, 0, \, 0  \rangle_B$. This means that we
do not need to introduce  the  $\theta$-vacua in  the CSM.

\begin{figure}[htbp] 
\vspace*{13pt}
\centerline{\psfig{file=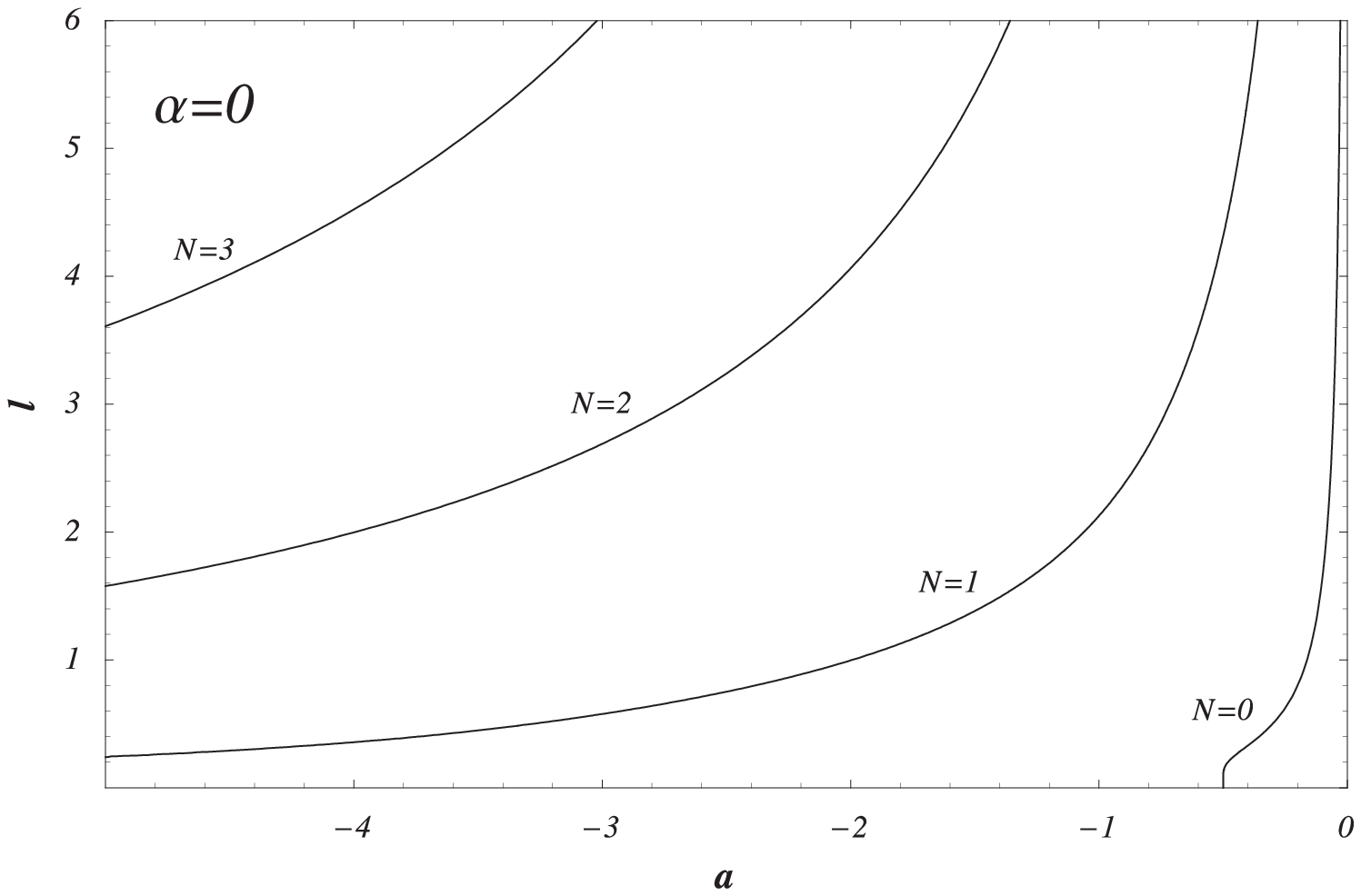}} 
\vspace*{13pt}
\fcaption{The
numerical solution for the parameter $a_{0,N}(l)$,
for $N=0, 1, 2, 3$ and a given value of $l$. The energies are
${\cal E}_{0,N,0}=-(e/\pi^{1/2})\, a_{0,N}$.}
\end{figure}

\begin{figure}[htbp] 
\vspace*{13pt}
\centerline{\psfig{file=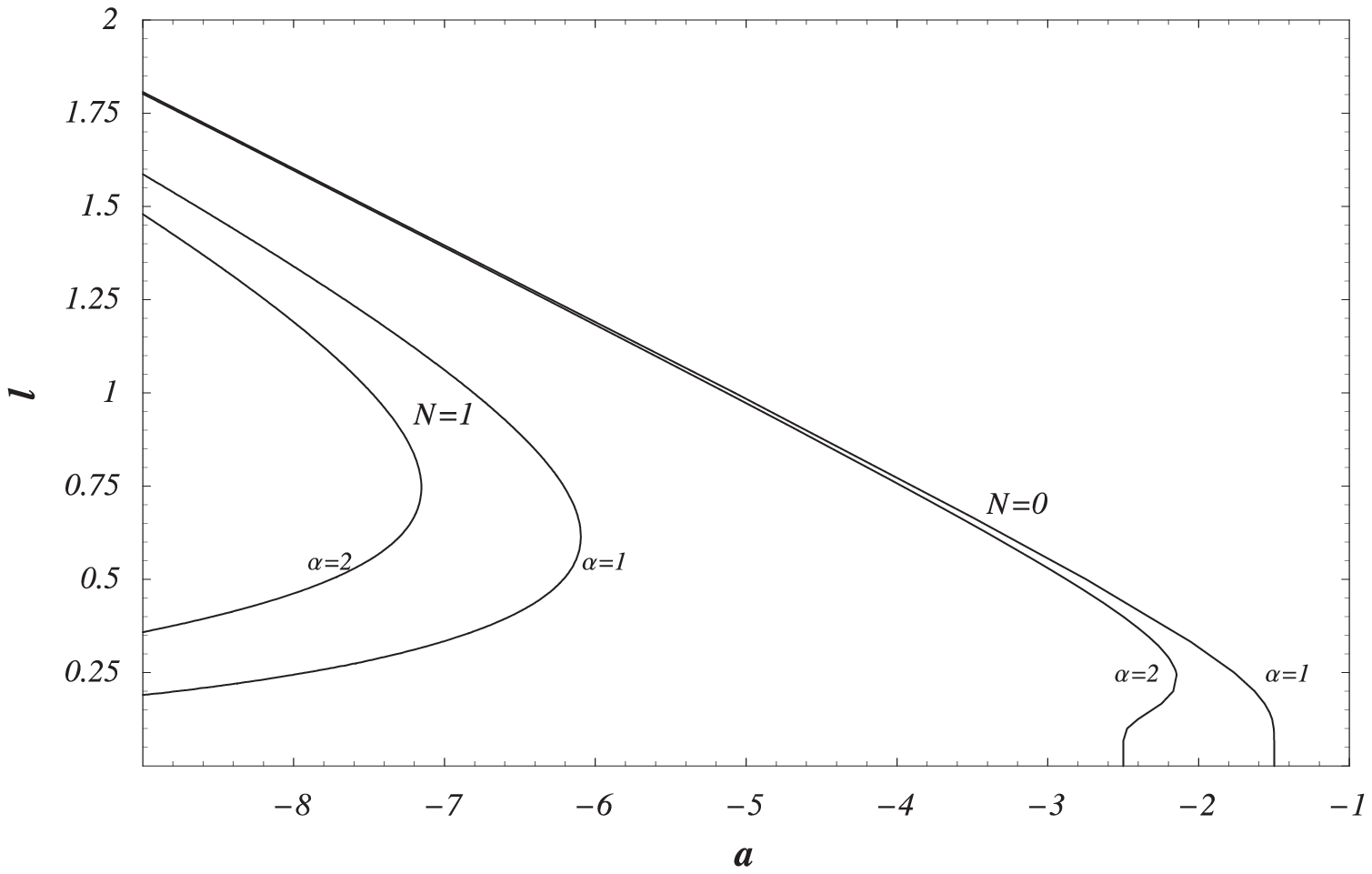}} 
\vspace*{13pt}
\fcaption{The
numerical solution of $a_{\alpha,N}(l)$, with $N=0,1$ and
$\alpha=1,2$, for a given value of $l$, is given.
The energies are
${\cal E}_{\alpha,N,0}=-(e/\pi{1/2})\, a_{\alpha,N}$.}
\end{figure}

The excited states
are obtained by applying  the creation
operators $({ j}_\pm{}^m)^\dagger$ to the zero modes constructed previously. Each
individual action raises the energy by ${\cal E}_m$, as can be seen from Eq.(\ref{HTOTAL}). The
excited states will be
labeled by
\begin{equation}
\label{STATES}
|\alpha,\ N, \ N_1,\ \dots, N_k, \dots \rangle_B,
\end{equation}
where $N_k$ is the the total number of times that the operators
$({ j}_\pm{}^k)^\dagger$ have
been applied to the corresponding minimun energy
state.
The total energy of the state (\ref{STATES}) is given by
\begin{equation}
{\cal E}_{\alpha,\, N, \,  N_1, \, N_2, \, \dots \, N_k,\,  \dots}= {\cal E}_{\alpha,\, N, \, 0}+
\sum_{k>0}
N_k \ {\cal E}_k.
\end{equation}

The fact that the chiral charge  $\bar Q_5=j_+{}^0-j_-{}^0 + e\,c\, L/\pi$ is conserved in the full
Hilbert space of the model deserves a separate discussion. We make the calculation
of the commutator $[\bar Q_5\, ,H_B ]$ in the Bogoliubov rotated frame, where $\bar Q_5$  preserves
the above expression and $H_B$ is given by  (\ref{HTOTAL}) . It is a direct matter
to verify  that  ${\bar Q}_5$ commutes with all the terms in the full Hamiltonian (\ref{HTOTAL}),
with the exception of the derivative
term. We analyze this piece in the sequel. Let us begin with
\be
\label{COMRP}
C_+{}^n= \left[\frac{\partial}{\partial\, c}, \, j_+{}^n\right]=
{\rm lim}_{s\rightarrow 0}
\sum_{m=-\infty}^{\infty} \frac{\partial}{\partial\, c}\left( \frac{1}{|\epsilon_n|^s}
\right)a^\dagger_{m} a_{m+n}
\ee
First, let us consider the action of $C_+{}^n, \, n\neq 0$ upon an arbitrary vector
\be
|\{m_i \}\rangle=\prod_i \, a_{m_i}{}^\dagger\, |0\rangle
\ee
in the positive-chirality fermionic Fock  subspace. In general, the subindex $m_i$ will take values
over an infinite subset of integer numbers. The only  non-zero result of the action of  the
$i^{th}$-term of (\ref{COMRP}) upon the above vector,  is to replace the $m_i +n$ fermion by the $m_i$
fermion, thus leading to a sum of linearly independent states. In this way, the $s\rightarrow 0$ limit
must be taken  separately in  each term of the series and no infinite summation occurs. Since
\be
\frac{\partial}{\partial\, c}\left( \frac{1}{|\epsilon_n|^s}\right)\approx \, - \, \frac{s}
{|\epsilon_n|^{s+1}},
\ee
this limit is zero and the operators commute.

Now, let us consider the $n=0$ case together with the action of $C_+{}^0$ upon the
local ground state $F_N(c) \times \left| {\varepsilon}_N,2N \right\rangle$ of each chirality sector. We
obtain
\ba
C_+{}^0\,|N\rangle_B&=& {\rm lim}_{s\rightarrow 0}
\sum_{m=-\infty}^{N-1} \frac{\partial}{\partial\, c}\left( \frac{1}{|\epsilon_n|^s}\right)\,
|N\rangle_B\nonumber \\
&=&
-\, \frac{e\, L}{2\, \pi}\, {\rm lim}_{s\rightarrow 0}\,  s\, \zeta(s+1, \, \frac{1}{2} +
\frac{e\,c\, L}{2\,\pi}-N)\, |N\rangle_B= -\, \frac{e\, L}{2\, \pi}\, |N\rangle_B,\nonumber \\
\ea
where $\zeta(s,\, q)$ is the standard Riemann  zeta-function. We have used the property
${\rm lim}_{s\rightarrow 0}\,  s\, \zeta(s+1, \, q)=1$
\cite{Riemann}. In analogous manner we obtain
\ba
C_-{}^0\,|N\rangle_B= \, \frac{e\, L}{2\, \pi}\, |N\rangle_B.
\ea
The above results lead to
\be
\left[\frac{\partial}{\partial\, c}, \, {\bar Q}_5  \right]\, |N\rangle_B= \left( C_+{}^0\,
- C_-{}^0\,  + \frac{e\, L}{\pi}\right) |N\rangle_B = 0.
\ee
Besides, any excited state is constructed by applying the raising operators $(j_\pm{}^n)^\dagger,
n\geq1$ to $|N\rangle_B$. These operators commute with ${\bar Q}_5$ and $\frac{\partial}{\partial\, c}$
in such way that the commutator $\left[\frac{\partial}{\partial\, c}, \, {\bar Q}_5  \right]$ is zero
in the full Hilbert space of the problem. This completes our proof that the fully gauge invariant
charge ${\bar Q}_5 $ commutes with the total
Hamiltonian (\ref{HTOTAL}).

Nevertheless, the axial-current anomaly is still present in the CSM, as we now
discuss. The charge
$Q_5$ arises from the current $
J_{5\, \mu}(x)={\bar \psi}(x)\gamma_\mu\, \gamma_5 \, \psi(x)$
which possesses the anomaly
\be
\label{SJ5AN}
\partial^\mu\, J_{5\, \mu}=-\, \frac{e}{\pi}\, E(x),
\ee
that can be directly calculated using the mode descomposition of $J_{5\, \mu}$, together with
the unrotated Hamiltonian
(\ref{HM}) and the Gauss law (\ref{AFORE}). On the other hand, one can  introduce the conserved
local  current
\be
\label{J5BAR}
{\bar J}_{5\, \mu}(x)=J_{5\, \mu}(x)-\frac{e}{\pi}\, \epsilon_{\mu\nu}\, A^\nu, \qquad
\epsilon_{01}=+1.
\ee
leading to the charge ${\bar Q}_5$.
Nevertheless, the current (\ref{J5BAR}) is not gauge invariant, so that  it cannot be restricted
to the physical Hilbert space of the problem.

Sumarizing this point, the axial current anomaly (\ref {SJ5AN}) is also present in the CSM, and it
cannot be removed, in spite that it is possible to define the conserved and gauge invariant modified
chiral charge
${\bar Q}_5$.

Finally we comment that our boundary conditions (\ref{BCWF})  are an unavoidable  consequence
of the compactification of the electromagnetic degree of freedom $c$.  These should be
contrasted with those apropriate for the NCSM ( Eqs. (3.15) of Ref. \cite{Manton}, or Eq. (48) of Ref.
\cite{Link}). The latter
are correctly designed to recover
the non-compact case, i.e. to go from the compact  gauge group $U(1)$ to the corresponding
universal covering. 
Also, for  a given $L\neq0$, the boundary conditions  for the CSM 
and those for  the NCSM can not be continuously connected between each other. This emphasizes  the fact that the
compactification condition (\ref{COMPACT}) has produced  a new version of the Schwinger model which is different
from the standard one.

\nonumsection{Acknowledgments}
\noindent
Partial support from the grants CONACyT  32431-E and DGAPA-UNAM-IN100397 is
acknowledged.

\nonumsection{References}
\noindent

\end{document}